\documentclass[final,5p,times,twocolumn]{elsarticle}

\usepackage{amssymb}

\journal{Physics Letters A}

\begin{document}

\begin{frontmatter}


\title{Electronic structure of a graphene superlattice with a modulated Fermi velocity}

\author{Jonas R. F. Lima}
\address{Instituto de Ciencia de Materiales de Madrid (CSIC) - Cantoblanco, Madrid 28049, Spain}
\ead{jonas.iasd@gmail.com}

\begin{abstract}

The electronic structure of a graphene superlattice composed by two periodic regions with different Fermi velocity, energy gap and electrostatic potential is investigated by using an effective Dirac-like Hamiltonian. It must be expected that the change of the Fermi velocity in one region of the graphene superlattice is equivalent to changing the width of this region keeping the Fermi velocity unchanged, provided that the time taken to charge carriers cross the region is the same. However, it is shown here that these two systems are not equivalent. We found extra Dirac points induced by the periodic potential and their location in the \textbf{k} space. It is shown that the Fermi velocity modulation breaks the symmetry between the electron and hole minibands and that it is possible to control the behavior of the extra Dirac points. The results obtained here can be used in the fabrication of graphene-based electronic devices. 

\end{abstract}


\end{frontmatter}

\section{Introduction}

Since its first successful experimental realization \cite{Novoselov}, graphene, an one-atom thick layer of carbon atoms arranged in a honeycomb lattice, has attracted a great deal of attention. Such interest is due to its unusual physical properties and potential application \cite{RevModPhys.81.109,RevModPhys.82.2673,RevModPhys.83.407}. The charge carriers in graphene, for instance, have an extremely high mobility and can be easily controlled by applying a gate voltage, which makes graphene a promising material for the fabrication of electronic devices. Controlling the energy band structure of graphene can help its application in electronics. It can be done, for instance, by applying and external periodic potential, \textit{i. e.}, a superlattice. 

The periodic potential structure in graphene can be generated by different methods, such as electrostatic potentials \cite{Bai,Barbier2,Park,Peeters,Tiwari,Wang2,Barbier,Wang,PhysRevB.86.205422}, magnetic barriers \cite{Ramezani,Sankalpa,Vasilopoulos,Luca} or the combination of both \cite{Zhai,Moldovan}. Despite the difficulty of fabricating graphene under nanoscale periodic potentials, it was already realized experimentally \cite{Marchini,Calleja,Sutter,Martoccia,Rusponi,Yan}. Although significant advances have been made in the understanding of how a periodic potential influences the electronic properties of graphene, very little further attention appears to have been given to the effects of a periodic modulation of the Fermi velocity in graphene superlattices \cite{Vasilopoulos2,Wang2013191,Nian,Lima1,Lima2}.

In our previous work \cite{Lima2}, we have found that it is possible to control the energy gap of graphene with Fermi velocity engineering when there is an energy gap modulation. It happens because the change of the Fermi velocity in a region of graphene is equivalent to changing the width of this region and keeping the Fermi velocity unchanged, provided that the time taken to charge carriers cross the region is the same. So, at the same way that it is possible to tune the energy gap of graphene controlling the width of two regions with different energy gap, it is possible to do the same changing the Fermi velocity. However, as will be shown here, this equivalence is related only with the energy gap, and does not remain for other electronic properties of graphene. In this paper we investigate the electronic structure of a graphene superlattice with a periodic modulation of the Fermi velocity and energy gap. Including the periodic potential, extra Dirac points appear. We show that the Fermi velocity affects the extra Dirac points in a different way of changing the widths of the regions,  being possible to  control the behavior of the extra Dirac points in different ways.

The paper is organized as follows. In Sec. 2 we solve the Dirac-like equation for graphne with a position dependent Fermi velocity, energy gap and electrostatic potential and find the dispersion relation. We also discuss how to obtain such a system experimentally. In Sec. 3 we analyze the electronic structure in two different cases. First we consider the Fermi velocity constant and investigate the electronic structure with a unequal well and barrier widths. Then, we consider a Fermi velocity modulation and an equal well and barrier widths. We compare these two cases and show that they are not equivalent. In Sec. 4 the paper is summarized and concluded.

\section{Model}

\begin{figure}[hpt]
\centering
\includegraphics[width=9cm,height=4cm]{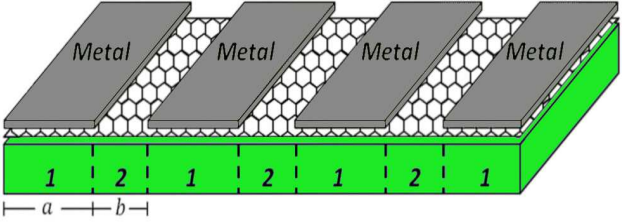}	
\caption{Schematic diagram of a graphene sheet deposited on a heterostructured substrate composed of two different materials which can open different energy gaps in different regions of the graphene sheet inducing a periodic modulation of the energy gap in graphene. Metallic planes are placed close to graphene sheet which induce a periodic velocity barrier. The period of the graphene superlattice is $a+b$.}\label{graphene}
\end{figure}

The effective two-dimentional Dirac Hamiltonian for a graphene superlattice with a position dependent energy gap and Fermi velocity is written as
\begin{eqnarray}
H=-i\hbar \left(\sqrt{v_F(x)}\sigma_x \partial_x  \sqrt{v_F(x)} +v_F(x)\sigma_y \partial_y\right) \nonumber \\
+V(x)\hat{1}+\Delta(x)\sigma_z ,
\end{eqnarray}
where $V(x)$ is an external periodic potential, $\Delta(x)$ is the energy gap, $\sigma_i$ are the Pauli matrices acting on the {\it pseudospin} related to the two graphene sublattices, $\hat{1}$ is the $2\times2$ unitary matrix and $v(x)$ is the Fermi velocity. The first term on the Hamiltonian above is modified in relation to the usual Dirac operator for graphene in order to have a Hermitian operator \cite{peres}. We are considering that the Fermi velocity, the energy gap and the potential change only in the $x$ direction. A schematic diagram is shown in Fig. \ref{graphene}. The graphene is deposited on a heterostructured substrate composed by two different materials, which can open different energy gaps in different regions of the graphene sheet that will be denoted by $\Delta_1=0$ and $\Delta_2=\Delta$. The modulation of the Fermi velocity can be obtained in graphene by placing metallic planes close to the graphene sheet, which will turn electron-electron interactions weaker and, consequently, modify the Fermi velocity \cite{Polini,Yuan}. The Fermi velocity in each region will be denoted by $v_1$ and $v_2$. The electrostatic potential for the two regions will be denoted by $V_1=0$ and $V_2=V$. The period of the system is $a+b$.

Following the calculations done in \cite{Lima2}, is straightforward to obtain that the dispersion relation is given by

\begin{eqnarray}
\cos (k_x l)&=&\cos(k_1a)\cos(k_2b) \nonumber \\
&+&\frac{k_y^2\hbar^2 v_1v_2+E(V-E)}{\hbar^2 v_1v_2k_1k_2}\sin(k_1a)\sin(k_2b) \; ,
\label{dispersion}
\end{eqnarray}
where $k_1=(E^2/\hbar^2 v_1^2 - k_y^2)^{1/2}$, $k_2=([(V-E)^2-\Delta^2]/\hbar^2 v_2^2 - k_y^2)^{1/2}$ and we have defined $l=a+b$. Note that at $V=\Delta=0$ and $v_1=v_2=v_F$ we recover the linear dispersion relation of a graphene sheet.

\section{Electronic Structure}

For the sake of comparison, in this section we will investigate the electronic structure of two special cases. First, we will consider a constant Fermi velocity and analyze the effects of an unequal well and barrier widths. Then, we will consider an equal well and barrier widths and see the influence of the Fermi velocity modulation on the electronic structure of the graphene superlattice. In what follows, we will consider a constant period of the graphene superlattice $a+b=l=60$ nm and we shall concentrate our discussion on the valence and conductance minibands only, assuming the Fermi level to be in between at any value of $V$.

\subsection{Unequal barrier and well widths}

\begin{figure}[hpt]
\centering
\includegraphics[width=8.2cm,height=4cm]{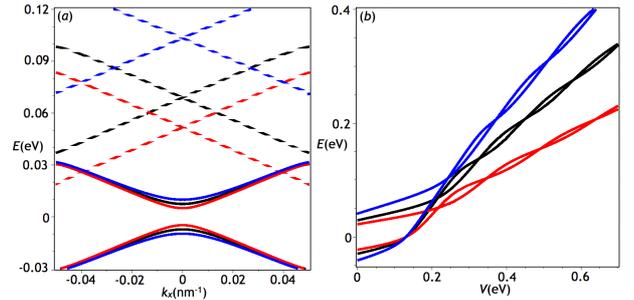}	
\caption{The dispersion relation (\ref{dispersion}) with $a=b=30$ nm (black), $a=20$ nm and $b=40$ nm (blue) and $a=40$ nm and $b=20$ nm (red) with $v_1 = v_2=10^6$ m/s, $\Delta=15$ meV and $k_y=0$. $(a)$ The electron and hole energies as a function of $k_x$, with $V=0$ (continuum lines) and $V=139.4$ meV (dashed black lines), $V=157.14$ meV (dashed blue lines) and $V=156$ meV (dashed red lines), revealing the possibility of closing the energy gap with the potential. $(b)$ The electron and hole minibands as a function of $V$ with $k_x=k_y=0$, which shows the oscillation of the energy gap.}
\label{kx2}
\end{figure}

In this section we will analyze the effects of an unequal well and barrier widths and we will consider $v_1 = v_2$. In Fig. \ref{kx2} $(a)$ the electron and hole minibands are plotted at $a=b=30$ nm (black), $a=20$ nm and $b=40$ nm (blue) and $a=40$ nm and $b=20$ nm (red) with $\Delta=15$ meV, $k_y=0$ and $V=0$ (continuum lines) and $V\neq 0$ (dashed lines). One can see that, at $V=0$, it is possible to tune the energy gap of graphene by changing the well and barrier widths. Increasing the potential the electron and hole minibands are shifted up, but not equally, which changes the energy minigap that may be zero. It is clear in Fig. \ref{kx2} $(b)$, where the electron and hole minibands are plotted as a function of $V$ with $k_x=k_y=0$. It is possible to see that the energy minigap oscillates and it is equal to zero at discrete values of $V$, which depends on the well and barrier widths. In the dashed lines in Fig. \ref{kx2} $(a)$ we consider the first value of $V$ that closes the minigap for the three cases. 

\begin{figure}[hpt]
\centering
\includegraphics[width=8.2cm,height=4cm]{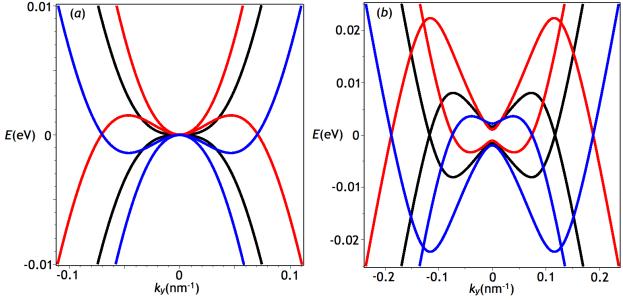}	
\caption{The dispersion relation (\ref{dispersion}) as a function of $k_y$, with $k_x = 0$. The colors here represent the same cases that in Fig. \ref{kx2}. We consider $(a)$ the same values of $V$ that in the dashed lines in Fig. \ref{kx2} and  $(b)$ $V=207.89$ meV (blabk lines), $V=234.12$ meV (blue lines) and $V=233.27$ meV (red lines).}
\label{ky2}
\end{figure}

In Fig. \ref{ky2} the electron and hole minibands are plotted as a function of $k_y$ for $k_x=0$. The colors represent the same cases that in Fig. \ref{kx2}. In order to be clear the influence of the periodic potential and the unequal well and barrier widths, the energies in this figure are counted from the contact point or from the minigap center for all values of $V$. In Fig. \ref{ky2} $(a)$ we consider the first value of $V$ that closes the minigap for the three cases. One can see that the contact point in $k_y$ direction has a parabolic behavior, while in the $k_x$ direction it is conical. So, the minibands have a lens-like shape. It can be seen also that when the well and barrier widths are equal, the electron and hole minibands are symmetric related to $E=0$. However, this symmetry is broken when the well and barrier widths are different.

In Fig. \ref{ky2} $(b)$ we consider $V=207.89$ meV (black), $V=234.12$ meV (blue) and $V=233.27$ meV (red), which are values between the first and second value of $V$ that close the gap at $k_y=0$. One can see that even though an energy minigap opens at $k_y=0$, extra Dirac points appear at $k_y\neq0$. These extra Dirac points appear when $V$ exceeds a critical value $V_c$, which is the first value that closes the minigap, and never disappear. So, from $V=V_c$, the graphene becomes gapless. One can note that the extra Dirac points that appears at $k_y\neq 0$ are not at the Fermi level when $a\neq b$. When $a>b$ ($a<b$) the Dirac points are shifted up (down) the Fermi level. 

In order to localize the contact points in \textbf{k} space let us take advantage of the implicit function theorem. One can conclude that at the contact points, where there is an intersection of the bands, the gradient (Jacobian) of the dispersion relation (\ref{dispersion}) should be zero. It happens only if $\sin k_1a=\sin k_2b=0$ and $\cos k_1a=\cos k_2b=\pm1$. So, one can write
\begin{equation}
k_1 a = \left(\frac{E^2}{\hbar^2 v_F^2} - k_y^2\right)^{\frac{1}{2}}a = n\pi 
\label{ka}
\end{equation}
and
\begin{equation}
k_2 b =\left(\frac{(V-E)^2-\Delta^2}{\hbar^2 v_F^2} - k_y^2\right)^{\frac{1}{2}}b = n\pi \; ,
\label{kb}
\end{equation}
where $n$ is an integer different from zero. Subtracting (\ref{kb}) from (\ref{ka}), one gets
\begin{equation}
E=E_{n}=\frac{V^2-\Delta^2}{2V}+\frac{\pi^2 n^2\hbar^2 v_F^2}{2V}\left(\frac{1}{a^2}-\frac{1}{b^2}\right) \; ,
\label{En}
\end{equation}
which gives the values of $E$ where there are contact points. Replacing the equation above in Eq. (\ref{ka}) one obtains
\begin{equation}
k_{y_n}=\pm \sqrt{\frac{E^2_{n}-\Delta^2}{\hbar^2 v_F^2}-\frac{n^2\pi^2}{a^2}} \; .
\label{m}
\end{equation}
As all contact points are located at $k_x=0$, with the equations above we have the complete location of the contact points. From the zeros of Eq. (\ref{m}) one obtains,
\begin{equation}
V_n=\frac{n\pi \hbar v_F}{a}+\sqrt{\left(\frac{n\pi \hbar v}{b}\right)^2+\Delta^2} \; ,
\label{Vn}
\end{equation}
which are the values of $V$ that close the minigap at $k_y=0$. The critical value of $V$ is $V_c=V_1$.

In order to be clear the location of the contact points it is important to explain that when $V$ exceeds $V_1$, two extra Dirac points appear at $k_y\neq 0$. When $V=V_2$ the energy minigap at $k_y=0$ closes and the system has now three contact points. When $V$ exceeds $V_2$ the minigap at $k_y=0$ opens but two extra Dirac points appear, leaving the system with four contact points and so on. When $V=V_n$ or $V_n < V < V_{n+1}$, $E_n$ and $k_{y_n}$ give the location of the contact point nearest to $k_y=0$ whereas $E_1$ and $k_{y_1}$ give the location of the contact point farthest to $k_y=0$. So, the location of the contact points are

\begin{equation}
(E, k_x, k_y)=(E_n, 0, k_{y_n}),(E_{n-1}, 0, k_{y_{n-1}}),..., (E_1, 0, k_{y_1}) \; .
\end{equation}

\subsection{Periodic velocity barrier}

\begin{figure}[hpt]
\centering
\includegraphics[width=8.2cm,height=4cm]{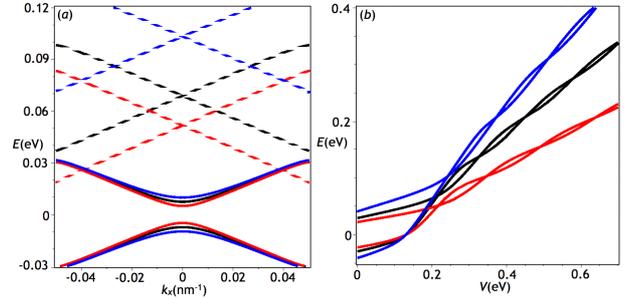}	
\caption{The dispersion relation (\ref{dispersion}) with $v_1=v_2=10^6$ m/s (black), $v_1=1.5 \times 10^6$ m/s and $v_2=0.75 \times 10^6$ m/s (blue) and $v_1=0.75 \times 10^6$ m/s and $v_2=1.5 \times 10^6$ m/s (red) with $a = b=30$ nm, $\Delta=15$ meV and $k_y=0$. $(a)$ The electron and hole energies as a function of $k_x$, with $V=0$ (continuum lines) and the dashed lines have the same values of $V$ that in Fig. \ref{kx2}. $(b)$ The electron and hole minibands as a function of $V$ with $k_x=k_y=0$, which shows the oscillation of the energy gap.}
\label{kx1}
\end{figure}

In this section we will consider a periodic velocity barrier with $a=b=30$ nm. In Fig. \ref{kx1} we consider $v_1=v_2=10^6$ m/s (black), $v_1=1.5 \times 10^6$ m/s and $v_2=0.75 \times 10^6$ m/s (blue) and $v_1=0.75 \times 10^6$ m/s and $v_2=1.5 \times 10^6$ m/s (red) with $\Delta=15$ meV and $k_y=0$. We recover the result obtained in \citep{Lima2}, which showed that it is possible to tune the energy gap of graphene by Fermi velocity engineering. One can note that the Fig. \ref{kx1} is exactly the same as Fig. \ref{kx2}. It happens because the values for the Fermi velocity were chosen here in such a way that the time taken to charge carriers cross the regions of the graphene superlattice is the same that in the case shown in Fig. \ref{kx2}. So, if we consider an one-dimensional case, the periodic velocity barrier is equivalent to the unequal well and barrier widths problem.

\begin{figure}[hpt]
\centering
\includegraphics[width=8.2cm,height=4cm]{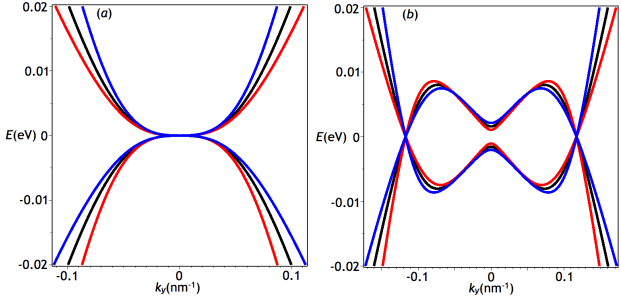}	
\caption{The dispersion relation (\ref{dispersion}) as a function of $k_y$, with $k_x = 0$. The colors here represent the same cases that in Fig. \ref{kx1}. We consider $(a)$ the same values of $V$ that in the dashed lines in Fig. \ref{kx2} and  $(b)$ the same values of $V$ that in Fig. \ref{ky2} $(b)$.}
\label{ky}
\end{figure}

However, with $k_y\neq0$ the two cases discussed here are not equivalent. In Fig. \ref{ky} we plotted the electron and hole minibands as a function of $k_y$ with $k_x=0$, where the colors represent the same cases that in Fig. \ref{kx1}. In Fig. \ref{ky} $(a)$ we consider the first value of $V$ that closes the gap. One can see that when $v_1\neq v_2$ the electron and hole minibands are asymmetric, one becoming narrower than the other. When one exceeds this value of $V$, an energy minigap opens at $k_y=0$ and extra Dirac points appear at $k_y\neq 0$, as shown in Fig. \ref{ky} $(b)$. But, in contrast to the extra Dirac points that appears in the case analyzed in the last section, the extra Dirac points here are all located at the Fermi level. Thus, the periodic Fermi velocity barrier can not shift up or down the extra Dirac points, as can be done by changing the well and barrier widths.

The location of the extra contact points can be found at the same way as in the last section. One can note that $k_1 = k_2$ when
\begin{equation}
E=E_0=\frac{Vv_1^2-v_1\sqrt{V^2v_2^2-\Delta^2v_2^2+\Delta^2v_1^2}}{v_1^2-v_2^2} .
\end{equation}
As all contact points are located at $k_x=0$, one can see that Eq. (\ref{dispersion}) with $a=b$, $k_x=0$ and $E=E_0$ is given by
\begin{equation}
1=\cos^2(k_1a)+\frac{k_y^2\hbar^2 v_1v_2+E_0(V-E_0)}{\hbar^2 v_1v_2k_1^2}\sin^2(k_1a) \; ,
\label{e0}
\end{equation}
which is satisfied when $k_1=m\pi/a$, where $m$ is an integer different of zero, because $m=0$ implies $k_1=0$, which makes the denominator $\hbar^2 v_1v_2k_1^2$ in Eq. (\ref{e0}) vanishes. This condition leads to
\begin{equation}
k_y =k_{y_m}=\sqrt{\frac{E_0^2}{\hbar^2v_1^2}-\left(\frac{m\pi}{a}\right)^2} \; ,
\label{kyn}
\end{equation}
which gives the values of $k_y$ where the contact points are located. As it was already observed, in contrast to the case discussed in the previous section, in this case for a given value of $V$ all contact points are located at the same value of energy $E_0$. The zeros of the equation above give the contact points at $k_y=0$. So, one can write
\begin{equation}
V_m=\frac{m\pi \hbar v_1+\sqrt{m^2\pi^2\hbar^2v_2^2+a^2\Delta^2}}{a},
\end{equation}
where the critical value of $V$ wherewith the graphene superlattice becomes gapless is $V_c=V_1$.

\section{Conclusions}

We analyzed the electronic structure of a graphene superlattice with a piecewise constant periodic potential, energy gap and Fermi velocity. We consider that the periodic energy gap is generated by an appropriate substrate composed by two different materials that open different energy gaps in different regions of the graphene and the Fermi velocity is modulated by metallic planes placed close to the graphene sheet. 

We compared two cases, one with an unequal well and barrier widths keeping the Fermi velocity constant and other with a periodic Fermi velocity barrier keeping the well and barrier widths equal. It was shown that with $k_y=0$ the two cases are equivalent, being possible to tune the energy gap of the graphene superlattice by changing the well and barrier widths or by Fermi velocity modulation. It was also found that the energy gap oscillates when the potential increases continuously. However, with $k_y\neq 0$ the two cases considered here are not equivalent. We found that extra Dirac points appears at $k_y\neq 0$ when $V$ exceeds the first value $V_1$ that closes the energy gap at $k_y=0$. As a consequence, the graphene superlattice becomes gapless after $V$ reaches a critical value $V_c=V_1$. We showed that in both cases the electron and hole minibands become asymmetric. But the extra Dirac points for the case of unequal well and barrier widths are shifted up or down the Fermi level, which does not happen with a periodic Fermi velocity barrier. The results showed here provide different ways of controlling the behavior of the extra Dirac points in graphene and can be used in the fabrication of graphene-based devices.

{\bf Acknowledgements}: This work was partially supported by CNPq and CNPq-MICINN binational.


\end{document}